\documentstyle[sprocl,epsfig]{article}

\newcommand{\Dirac}{\rlap {\hspace{-0.5mm} \slash} D}

\def\be{\begin{eqnarray}}
\def\ee{\end{eqnarray}}
\def\bea{\begin{array}}
\def\eea{\end{array}}
\def\bei{\begin{itemize}}
\def\eei{\end{itemize}}
\begin{document}

\title{THE INFRARED LIMIT OF THE QCD DIRAC SPECTRUM} 
\author{J.J.M.  Verbaarschot}

\address{Department of Physics and Astronomy, SUNY Stony Brook, \\
Stony Brook, NY 11790, USA,\\
e-mail: verbaarschot@nuclear.physics.sunysb.edu}

\maketitle\abstracts{The distribution of the low-lying QCD Dirac spectrum
is analyzed by means of partial quenched chiral perturbation theory. 
We identify an
energy scale below which the valence quark mass dependence of the QCD
partition function is given by chiral Random Matrix Theory. This is the
the domain where the low-energy QCD partition function is dominated by the
fluctuations of the mass term of the Goldstone bosons. 
In mesoscopic physics this domain is known
the ergodic domain and the corresponding energy scale is the Thouless
energy. 
}

\section{Introduction}
It has become clear, both from numerous lattice QCD simulations 
\cite{DeTar,Karsch}
and hadronic phenomenology, that QCD at low energies is characterized by
chiral symmetry breaking and confinement. Because of spontaneous breaking
of chiral symmetry, the low-lying excitations are given by the associated
Goldstone modes. Because of confinement, these are the only low-lying 
excitations. In this lecture, we wish to point out another important 
feature of the low energy QCD partition function, namely disorder. Ultimately
we would like to answer the question whether confinement can occur if the
classical motion of quarks in generic Yang-Mills fields is not chaotic?

One way to understand confinement is by analogy with  impurity scattering
\cite{Polonyi}. The argument starts from the 
 semiclassical expression for the quark propagator in an  external 
gauge field  
given by a sum over classical trajectories weighted by a phase. 
Its average over all gauge fields  vanishes because of this phase. 
On the other hand, 
the pion propagator given by the absolute value of the quark
propagator survives averaging over all gauge field configurations.
We will analyze the quark propagator using a spectral representation
of the Dirac operator. From the study of disordered systems we know that
the eigenvalues of localized states with and exponentially falling
wave function are uncorrelated, whereas 
 eigenvalues of extended states with
a power-like fall-off of the wavefunction are correlated according to
Random Matrix theory. Spectral correlations therefore constitute 
 a sensitive measure for the tail of the wave functions. 
One of the questions we are interested in is 
 whether the correlations of the Dirac
eigenvalues support this picture of confinement.

The spectrum of the Dirac operator  and the breaking chiral
symmetry are not unrelated. According to the Banks-Casher formula 
\cite{BC} the 
order parameter of the chiral phase transition is given by
\be
\Sigma = \frac{\pi\rho(0)} V,
\label{Banks}
\ee
where $\rho(\lambda) $ is the average spectral density
$\rho(\lambda) = \langle\sum_k \delta(\lambda -\lambda_k) \rangle,
$
and $V$ is the volume of space time. The average $\langle \cdots
\rangle$ is over the QCD action.
With the smallest eigenvalues of the Dirac operator are spaced
$\Delta \lambda =  \pi/{\Sigma V},$
it is natural to study this part of the spectrum by introducing the
microscopic variable $u = V\lambda \Sigma$, and the microscopic
spectral density defined by \cite{SVR}
\be
\rho_s(u) = \lim_{V\rightarrow \infty} \frac 1{V\Sigma} \langle
\rho(\frac u{V\Sigma})\rangle.
\label{rhosu}
\ee

We expect that this limit converges to a universal function
determined by the global symmetries of the QCD Dirac operator.
Using partially quenched chiral QCD partition function we will show
that $\rho_s(u)$ is given by chiral Random Matrix Theory  (chRMT) to be
discussed in the next section. We will find
the domain of validity of chRMT and will calculate the Dirac spectrum
beyond this  domain. An extensive discussion of the material of
this talk with  a complete list of references has appeared recently
\cite{Kyoto}, and several excellent recent reviews on Random Matrix Theory 
and disorder are available \cite{HDgang,Montambaux}. 

\section{Chiral Random Matrix Theory}

The chiral random matrix partition function
 with the global symmetries of the QCD partition function 
is defined by {\cite{SVR,V}}
\be
Z^\nu_\beta({\cal M}) =
\int DW \prod_{f= 1}^{N_f} \det{\left (\begin{array}{cc} m_f & iW\\
iW^\dagger & m_f \end{array} \right )}
e^{-\frac{N \beta}4 \Sigma{\rm Tr}W^\dagger W },
\label{ranpart}
\ee
where $W$ is a $n\times m$ matrix with $\nu = |n-m|$ and
$N= n+m$.
As is the case in QCD, we assume that the equivalent of the topological charge
$\nu$ does not exceed $\sqrt N$,
so that, to a good approximation, $n = N/2$.
Then the parameter $\Sigma$ can be identified as the chiral condensate and
$N$ as the dimensionless volume of space time (Our units are defined
such that the density of the modes $N/V=1$).
The matrix elements of $W$ are either real ($\beta = 1$, chiral
Gaussian Orthogonal Ensemble (chGOE)), complex
($\beta = 2$, chiral Gaussian Unitary Ensemble (chGUE)),
or quaternion real ($\beta = 4$, chiral Gaussian Symplectic Ensemble (chGSE)).
For QCD with three or more colors and quarks in the fundamental representation
the matrix elements of the Dirac operator are complex and we have $\beta = 2$.
The ensembles with $\beta =1 $ and $\beta =4$ are relevant in the case
of two colors or adjoint fermions. 
The reason for choosing a Gaussian distribution of the matrix elements is
its mathematically simplicity. It can be shown that the correlations
of the eigenvalues on the scale of the level spacing do not depend on
the details of the probability distribution 
\cite{Brezin,ADMN,AkemannFV,Sener1,GWu,Seneru,Kanzieper}.

\section{Scales in the Dirac Spectrum}

Of course, QCD is not chiral Random Matrix Theory. This
implies the existence of a scale beyond which the correlations 
QCD Dirac spectra are no longer given by chRMT. 
We have studied such scale by means of 
the valence quark mass dependence of the
chiral condensate \cite{Vplb,Osbornprl,OTV,DOTV,TV}  defined by
\cite{Christ}
\be
\Sigma(m_v) = \langle {\rm Tr} \frac 1 {m_v + D} \rangle.
\ee
Here, the Euclidean Dirac operator is denoted by $D$ and the average
is over the Euclidean QCD action which includes a fermion determinant
that depends on the the sea-quark masses. The spectrum of $D$ is
directly related to to $\Sigma(m_v)$,
\be
\rho(\lambda)/V = \frac 1{2\pi} \left (\Sigma(i\lambda +\epsilon) -
\Sigma(i\lambda -\epsilon) \right ).
\ee
 As is the case in standard chiral
perturbation theory \cite{GL,LS}, an important scale 
is the mass where the Compton wave
length of  Goldstone bosons corresponding to the valence quark masses
is equal to the linear size of the system \cite {Vplb}, i.e. the 
mass scale $m_c$ given by 
\be
\frac {2m_c \Sigma}{F^2} = \frac 1{L^2}.
\label{scale}
\ee
 We have identified the range \cite{Vplb,Osbornprl,OTV,DOTV,TV} 
\be
m_v \ll m_c = \frac {F^2}{\Sigma L^2}. 
\ee
as the domain of validity of chRMT.
In other words, below the scale $m_c$ the correlations of the QCD
Dirac eigenvalues are completely determined by the global symmetries
of the QCD partition function. 

A second important scale is the average position of the smallest 
eigenvalue of the Dirac operator, which is approximately given by
the average spacing of the eigenvalues near $\lambda = 0$. With
spacing given by $1/\rho(0)$ and the Banks-Casher formula (\ref{Banks})
we find
\be
\lambda_{\rm min} = \frac {\pi}{\Sigma V} .
\ee
In the QCD Dirac spectrum we can thus distinguish \cite{Vplb} three
different scales,
\be 
\lambda_{\rm min} \ll m_c  \ll \Lambda_{QCD}.
\ee
These scales separate four different domains. Similar domains occur in
mesoscopic physics \cite{Altshuler}. 
The reason is that the transport of electrons in
disordered samples is described by Goldstone modes.
The classical
propagation of Goldstone modes is given by  a diffusion
equation \cite{Stone} and therefore $\hbar/m_c$ can be interpreted as
the  time  for an initially
localized wave packet to diffuse over the length  of the sample
(see for example the book by Efetov \cite{Efetovbook}). The
domain $m_v \ll m_c$ is therefore known as the ergodic domain.
In mesoscopic physics  the
scale corresponding to  $m_c$ is known as the Thouless energy.

\section{Partially Quenched Chiral Perturbation Theory}

The scales in QCD Dirac spectra can be studied more precisely via
the generating function of the chiral condensate. This is the
partially quenched QCD partition function 
which, in addition to the usual quarks describing the
fermion determinants, contains additional quarks
and their bosonic super-partners which describe the quenched valence 
quark mass dependence of the chiral condensate.  It is defined by
 \cite{Morel,pqChPT,GolLeung}
\be
Z^{\rm pq}(m_v,J)  ~=~ \int\! [dA] 
~\frac{\det(\Dirac +m_v+J)}{\det(\Dirac +m_v)}\prod_{f=1}^{N_{f}}
\det(\Dirac + m_s) ~e^{-S_{YM}[A]} ~,
\label{pqQCDpf}
\ee
where $\Dirac$ is the Euclidean Dirac operator and $S_{YM}[A]$ is the
Euclidean Yang-Mills action.
The valence quark mass dependence of the chiral condensate is then given
by \cite{Christ,Vplb}
\be
\Sigma(m_v) = \frac 1V\langle \sum_k \frac 1{m_v +i\lambda_k} \rangle  =
\frac 1V \left . \partial_J \right |_{J=0} \log Z^{\rm pq} (m_v, J).
\ee
 For
a confining theory such as QCD, the low energy excitations of this
theory are the 
Goldstone modes associated with the spontaneous breaking of chiral
symmetry. To lowest order in the momenta, the effective action of these
modes is completely determined by Lorentz invariance and chiral symmetry
breaking. In agreement with the Vafa-Witten \cite{Vafa} theorem and maximum
breaking of chiral symmetry \cite{Shifman-three} 
we expect that the chiral symmetry
is broken according to \cite{OTV,DOTV}
\be
Gl(N_f+1|1) \times Gl(N_f+1|1)\rightarrow Gl(N_f+1|1) .
\ee
In order to obtain convergent integrals, the
effective partition function is then based on the maximum Riemannian
submanifold of the symmetric superspace $Gl(N_f+1|1)$ which will
be denoted by $\hat Gl(N_f+1|1)$. It
contains 
a kinetic term determined by Lorentz invariance and chiral symmetry
and a mass terms determined by the pattern of chiral symmetry
breaking. An explicit axial $Gl(1|1)$ symmetry breaking  
 is included as well
\be
Z(\theta,\hat {\cal M}) &=& \int_{U\in \hat 
Gl(N_f+1|1)} dU \exp \int d^4 x \left [
\frac{F^2}{4} \;
{\rm Str} (\partial_\mu U \; \partial_\mu U^{-1})
\right .\nonumber \\
&+& \left . \frac{\Sigma}{2} \; {\rm Str} (\hat{\cal M} U+\hat{\cal M} U^{-1})
+\frac{F^2 m_0^2}{12} \; (\frac {\sqrt 2 \Phi_0}F-\theta)^2\right ].
\label{zthetam}
\ee

In terms of 
the action of this  partition function, 
the scale $m_c$ is the scale
where the fluctuations of the mass term and the kinetic term
are of equal order of magnitude. For valence quark masses 
$m_v \ll m_c$ the fluctuations of the 
kinetic term can be neglected and the calculation
of the valence quark mass dependence of the chiral condensate is
reduced to the calculation of zero dimensional super-integrals. Projecting
onto fixed  topological charge by Fourier decomposing the 
$\theta$-dependence  the zero-dimensional effective partially
quenched partition
function is given by\cite{OTV}
\be
Z^\nu_{\rm eff}(\hat{\cal M}) =
\int_{U\in \hat Gl(N_f+1|1)} dU \,{\rm Sdet}^\nu U \, e^{
 V\frac{\Sigma}{2} \; {\rm Str} (\hat{\cal M} U+\hat{\cal
M} U^{-1})}.
\label{superpart}
\ee
As an illustration, we mention the parameterization of the integration
manifold for $N_f = 0$. Then $U \in \hat Gl(1|1)$ is given by
\be
U = \left ( \begin{array}{cc} e^{i\phi} & \alpha \\ \beta & e^s
\end{array} \right ),
\ee
where $\phi \in [0,2\pi]$, $s \in \langle -\infty, \infty \rangle$, and 
$\alpha$ and $\beta$ are Grassmann variables.
In terms of these variables
the integration measure is
simply given by $dU = d\phi d s d\alpha d\beta$. The calculation of the
integrals is 
straightforward in this case.
For arbitrary $N_f$ and $\nu$, on the other hand, 
the integrals in (\ref{superpart}) are technically quite
involved, but we have succeeded \cite{OTV,DOTV} 
to obtain an analytical result for
the valence quark mass dependence of the chiral condensate,
%To give you some of the flavor of the calculations we only give
%the parameterization the integration manifold of the simplest case
%for $N_f = 0$,
%\be
%U =\exp {\left ( \begin{array}{cc} 0 &\alpha \\ \beta & 0 \end{array} \right )}
% \left ( \begin{array}{cc} e^{i\phi} &0 \\ 0 & e^s \end{array} \right ).
%\ee
%The calculation involves the derivation of the integration measure and
%the evaluation of the superintegrals. 
%By an explicit
%calculation of the superintegrals we have shown that in this domain
%the valence quark mass dependence of the chiral condensate is given by
%chiral Random Matrix Theory with the result

\be
\frac {\Sigma(x)}{\Sigma} = x(I_{a}(x)K_{a}(x)
+I_{a+1}(x)K_{a-1}(x)),
\label{val}
\ee
where $a = N_f+|\nu|$.  It
coincides with the result derived from chRMT.
The microscopic spectral density follows from
the discontinuity and agrees with the chRMT result \cite{VZ,V} as well.

Our effective partition function is amenable to 
chiral Perturbation Theory allowing us to go beyond the Random Matrix
Theory limit of QCD.
Taking into account the
kinetic term to one-loop order  we have obtained an exact analytical
expression for the  QCD Dirac
spectrum in the domain where chiral perturbation theory applies, i.e.
for $m_v \ll \Lambda_{QCD}$.
Among others, we have calculated the slope
of the Dirac spectrum and shown that the spectral density diverges
logarithmically in the quenched limit. The slope vanishes for two
massless flavors, not only for the standard pattern of chiral symmetry
breaking \cite{SS}, 
but also for other patterns of chiral symmetry breaking
relevant for QCD with 
two colors and fundamental fermions or QCD with adjoint fermions \cite{TV}.

The microscopic spectral density near zero virtuality was first studied
in terms of the valence quark mass dependence of the chiral condensate.
The lattice data for $\Sigma(m_v)$ were obtained by the Columbia
group \cite{Christ}
for  two dynamical flavors with  sea-quark mass $ma = 0.01$ and $N_c= 3$ on a
$16^3 \times 4$ lattice. 
In Fig. 4 we plot \cite{Vplb} 
the ratio $\Sigma(m_v)/\Sigma$ as a function
of 
$x=m_v V \Sigma$ (the 'volume' $V$ is equal
to the total number of Dirac eigenvalues), and compare the results
with the universal curves obtained from pq-QCD and chRMT (see eq. (\ref{val})).
 We observe
that the lattice data for different values of $\beta$ fall on a single curve.
Moreover, in the mesoscopic range
this curve coincides with the random matrix prediction for $N_f = \nu = 0$.
\begin{center} 
\begin{figure}[!ht]
\centering\includegraphics[width=55mm,angle=270]{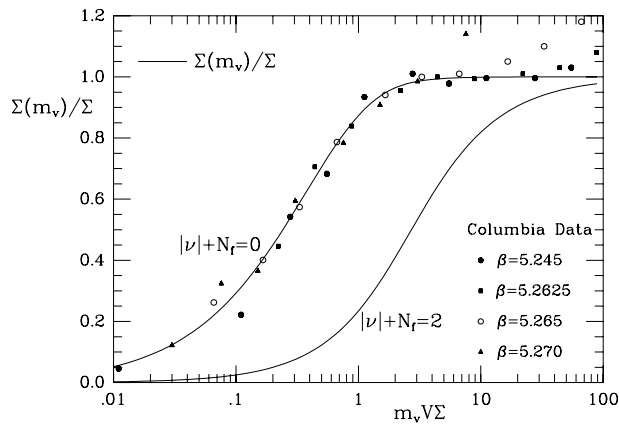}
%{sys$scratchdisk:[verbaarschot.zak]valzak.ps}
\caption[]{The valence quark mass dependence of the chiral condensate
$\Sigma(m)$ plotted as $\Sigma(m_v)/\Sigma$ versus $m_v V\Sigma$. 
The dots and
squares represent lattice results by the Columbia group \cite{Christ}
 for values of $\beta$ as indicated in the label of the figure.}
\label{fig5}
\end{figure}
\end{center}
Of course this is no surprise. For quark masses  much less than
the current quark mass, the fermion determinant has no bearing on the Dirac
spectrum and we have effectively $N_f =0$.
At finite lattice spacing, the Kogut-Susskind action does not have
the axial $U(1)$ symmetry and  the effects of topology become  only visible 
in the continuum limit.
We observe a plateau in the valence quark mass dependence of the
chiral condensate. The increase at larger masses is also found in
partially quenched chiral perturbation theory and is due to the
effects of the kinetic term. The scale $m_c$ (see eq. (\ref{scale}))
is roughly located at 
the center of the plateau. Meanwhile, the valence quark mass dependence
of the chiral condensate and the scalar susceptibility
have also been investigated for $SU(2)$ gauge
field configurations and different types of fermions 
\cite{Tiloval,Poulval,Karlval}. In all cases agreement between lattice
QCD spectra and the universal result (\ref{val}) has been found.
A detailed analysis
of scalar susceptibilities in terms of quenched
chiral perturbation theory was recently performed by Berbenni-Bitch et al.
\cite{Tiloval}.
The microscopic spectral density can be calculated directly
and a similar type of agreement with chRMT has been observed
\cite{Vinst,Tilo,DamSU3,TiloSU3,Heller,Kiskis,Hip,Tilo99,Damtopo,hiplat99}.
\begin{figure}[!ht]
\centering
\includegraphics[width=70mm,height=51mm]{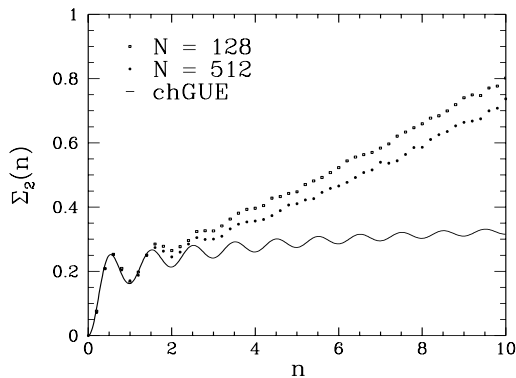}
\caption{The number variance $\Sigma_2(n)$ versus $n$
approximation for an interval starting at $\lambda = 0$. The total number
of instantons is denoted by $N$.}
\end{figure}

Another interesting observable in the number variance,
$\Sigma_2(n)$, defined as the
variance of the number of eigenvalues in a interval containing 
$n$ eigenvalues on average.
It  can be obtained in partially quenched chiral
perturbation theory from the double discontinuity of the pseudo-scalar
susceptibility \cite{DOTV} (A recent discussion of the
relation between the pion propagator and the number variance can be
found in \cite{Carter}). The number variance has been calculated
both  by means lattice QCD \cite{many,Guhr-Wilke}
 and instanton liquid~\cite{Osbornprl}
simulations. In Fig. 1, we show \cite{Osbornprl} 
$\Sigma_2(n)$ versus
$n$ for eigenvalues obtained
from the Dirac operator in the background of instanton liquid gauge
field configurations.
The chRMT result,  given by the solid curve, is reproduced
up to about two level spacings.
In units of the  average level spacing, $\Delta=1/\rho(0)=\pi/\Sigma V$,
the energy $E_c$ is given by
$n_c \equiv {E_c}/{\Delta} = {F^2 L^2}/\pi$.
For an instanton liquid with instanton density $N/V =1$ we find that
$n_c \approx 0.07 \sqrt N$. We conclude that chRMT appears to
describe the eigenvalue correlations up to the
predicted scale.
                               
\section{Conclusions}
We have found that for mass scales where chiral perturbation theory
is valid the fluctuations of the QCD Dirac eigenvalues can be obtained
analytically from the partially quenched effective chiral Lagrangian.
For mass scales below $F^2/\Sigma L^2$ the eigenvalue correlations
are given by chRMT. A possible semiclassical interpretation is that
quarks undergo a chaotic-diffusive motion in four Euclidean dimensions
plus one artificial time dimension. This observation is consistent with 
a picture of confinement based on impurity scattering.

\section{Acknowledgments} I wish to thank the organizers of the Trento
workshop and I gratefully acknowledge all my collaborators in this
project. This work was partially supported by the US DOE grant
DE-FG-88ER40388.

\end{document}